\documentstyle[12pt]{article}
\begin{document}
\begin{flushright}    
UCSBTH-96-07\\           
gr-qc/9604051\\       
\end{flushright}      
\begin{center}
{\bf
THE ORIGIN OF BLACK HOLE ENTROPY IN STRING THEORY\footnote{To appear in
the proceedings of the Pacific Conference on Gravitation and Cosmology,
Seoul, Korea, February 1-6, 1996.}}
\vskip 1cm
Gary T. HOROWITZ\footnote{e-mail: gary@cosmic.physics.ucsb.edu}
\vskip 1cm
{\it
Department of Physics, \\
University of California, \\
Santa Barbara, CA.  93106 USA
}
\end{center}
\vskip 1cm
\noindent {\bf Abstract}
\vskip 0.7cm
I review some recent work in which  the quantum
states of string theory which are associated with certain black holes
have been identified and counted. For large black holes, the number 
of states turns out to be precisely the exponential of the 
Bekenstein-Hawking entropy. This provides a statistical origin for black hole
thermodynamics in the context of a potential quantum theory of gravity.

\newcommand{\rp} {r_+^2}
\newcommand{\rmn} {r_-^2}
\newcommand{\be}{\begin{equation}}
\newcommand{\ee}{\end{equation}}
\newcommand{\lref}{\bibitem}
\newcommand{\NS}{${\cal NS }\ $}
\newcommand{\R}{${\cal R }\ $}

\vskip 1cm
\noindent {\bf I. Introduction}
\vskip 0.7cm

When I was first asked to give a title for this talk, two months before
the conference, I chose ``Gravitational insights from string theory".
By the time the conference began, I had changed my title to ``Can string
theory answer fundamental questions in gravitational physics?" As I finish
writing up this contribution for the proceedings,
now two months after the conference, my title has
changed again. I hope this illustrates, not my inability to stick to one
topic, but rather the rapid pace of progress in this field. This is an
exciting time in string theory. There have been major developments in
several different directions. New results have been appearing almost
every week. Since it is not possible to review all aspects of this work
here, I have chosen one topic that I believe is of most interest to the
community of gravitational physicists. Even with this restriction,
it is difficult
to review  such a rapidly evolving field, but I will try
to summarize our current understanding.

The early days of string theory were largely devoted to developing the
perturbative expansion of the theory. However, the recent progress  
has lead to an understanding of  
a number of nonperturbative effects.
With this progress, string theory is finally at a stage where it can
begin to answer fundamental questions such as:
What is the origin of black hole entropy?

This question goes back over twenty years to the early 1970's. At that
time it was noticed that there was a striking similarity between the laws
of black hole mechanics \cite{bch,bek} which include
\be
dM = {1\over 8\pi G} \kappa dA, \qquad \qquad \delta A \ge 0
\ee
(where $\kappa$ is the surface gravity and $A$ is the area of the 
event horizon)
and the laws of thermodynamics
\be 
dE = T dS, \qquad \qquad
\delta S \ge 0.
\ee
Hawking showed \cite{haw}
that this was more than an analogy with his famous discovery
that black holes indeed radiate a thermal spectrum with temperature
$T= \hbar \kappa/2\pi$. This implies that black holes have an entropy equal
to one quarter of their horizon area in Planck units:
$S_{BH} = A/4G\hbar$. 
Thus general relativity
and quantum field theory in curved spacetime give a beautiful thermodynamic
description of  black holes. But thermodynamics is only an approximation
to a more fundamental description based on statistical mechanics. The
challenge has been to find this more fundamental description, and find the
the $e^{S_{BH}}$ states associated with a black hole.
Since the black hole
temperature is a quantum mechanical effect (proportional to $\hbar$)
the statistical description must presumably involve a quantum theory
of gravity, which makes this problem particularly difficult.

String theory is a promising candidate for a  quantum theory of gravity,
and it has recently provided an answer to this question for a certain
class of black holes. (For some earlier work, see [4 - 17].)
Since there have been several strange black holes
discussed in the context of string theory, let me emphasize that we will
consider
ordinary four (or five) dimensional black holes and not black holes
in lower dimension, massless black holes, or black holes with zero
horizon area (although these latter objects will play a role at an 
intermediate stage in our discussion).

Before we begin, let me clarify a point which
may later cause confusion. As relativists, we are used to setting Newton's
constant $G$ equal to one, and measuring mass in units of length. In string
theory, there is another fundamental length scale $l_s$ called the string scale
which is set by the string tension
$T \sim 1/l_s^2$. Newton's constant
is related to  $l_s$ by $G \sim g^2 l_s^2$ (in four dimensions) where $g$ is
the string coupling constant. Although one could still set $G=1$, it is
often more convenient to set $l_s =1$  (as well as $\hbar =1$) since
masses of string states are multiples of $1/l_s$. The gravitational field
produced by these states depends on $G$ and increases with the string coupling.

Another reason for not setting $G$ to one is that Newton's constant 
is dimension dependent. If we consider five dimensional general relativity
and restrict to spacetimes with a translational
symmetry in a compact direction of length $L$ asymptotically, 
then it is natural to set
$G_5 = G_4 L$. This is because the four dimensional reduced action then
takes the standard
form
\be
{1\over 16 \pi G_5} \int d^5 x \sqrt{-g_5} R_5 = {1\over 16 \pi G_4}
  \int d^4 x \sqrt{-g_4} (R_4 + \cdots)
\ee
where the dots denote matter fields that arise from the dimensional reduction.
This is important for understanding the entropy of extended black holes.
Consider the five dimensional vacuum solution which is the product
of the four dimensional Schwarzschild metric
and a circle of length $L$. Since the
horizon area is proportional to $L$, one might think that the entropy is
also proportional to $L$. But the entropy is $A/4$ in Planck units, and
these units are different in four and five dimensions.
In fact
\be
{A_5 \over 4 G_5} = {A_4\over 4 G_4}
\ee
so the entropy is independent of $L$ can be calculated in either
dimension with the same result.

In the next section, I discuss in general terms the explanation of
black hole entropy provided by string theory. In section III, some
more details are provided by considering an explicit example. Section
IV contains further discussion of these recent results and some open
problems.

\vskip 1cm
\noindent {\bf II. Black Hole Entropy}
\vskip 0.7cm
To begin our discussion, let me briefly review some basic aspects
of string theory. One starts 
with the assumption that fundamentally, all particles
are excitations of a one dimensional extended object - a string.
When this string is quantized in flat spacetime, one finds 
an infinite tower of states. There are a finite number of
massless fields which include 
a scalar field (called the dilaton)
whose asymptotic value determines the string coupling $g$, gauge fields,
and a spin two field identified with the graviton.
Thus, one mode of the
string corresponds to a linearized perturbation of the metric.
For every integer $N$, there are  massive states with $M^2 = 
N/l_s^2$. These massive states have a large degeneracy.
The number of states with mass $M$ grows 
roughly like $e^M$. One might wonder whether this degeneracy
has something to do with black hole entropy, but at least naively
it does not seem to work\footnote{See \cite{sus} for a less naive interpretation
in which a resolution of this discrepancy is proposed.}:
The number of black hole states should grow like $e^{M^2}$ not $e^M$.
We will see that the explanation of black hole entropy is more subtle.

Strings interact by a simple splitting and
joining interaction so the two dimensional worldsheet remains smooth.
Classical graviton scattering computed this way agrees
with the perturbative expansion of general
relativity. But string theory is not just a theory in flat spacetime.
When strings are quantized in curved spacetime, there is a 
consistency condition which yields an equation of motion for the spacetime
metric and other massless fields, as well as a restriction on the
spacetime dimension. In the simplest  case,
this requires ten dimensions. 
The equation for the spacetime metric takes the form of Einstein's 
equation with matter consisting of the other massless fields in the theory,
plus an infinite series of higher order terms multiplied by powers of the
string length $l_s$.
The first part of the equation of motion (not multiplied by powers of $l_s$)
is called the `low energy string equation'.
When the curvature is
small compared to $1/l_s^2$ the higher order terms are negligible and
the theory reduces to general relativity coupled to certain matter fields.
When the curvature is of order $1/l_s^2$ or greater, the higher order
terms are important. In fact, in this case, the metric may not be 
well defined. This is because the theory does not prefer $g_{\mu\nu}$
over $g_{\mu\nu} + l_s^2 R_{\mu\nu} + ...$ where the dots denote arbitrary
second rank tensors constructed from the curvature and its derivatives.
Since $l_s$ is
roughly the size of a string, one can view this as saying that the
metric is only well defined on length scales larger than a string.

A key step in the recent developments was the realization that in
addition to the states described by string fluctuations, there are also
soliton states in the theory. Solitons are static, finite energy
classical solutions
to the field equations. In string theory there are several types of
solitons. Two are familiar from field theories; nonsingular magnetic monopoles,
and black holes. Since string
theory is naturally formulated in higher spacetime dimensions, there also exist 
extended monopoles and extended 
objects surrounded by horizons called black strings, black
membranes, and more generally, black $p$-branes for $p$ dimensional
extended objects.  Solutions to the low energy string equations of 
this type have been known for some time, and since the curvature is small
at the horizon for large black holes, these should be close to exact
string solutions. (For a recent review, see \cite{dkl}.) 
However,
except for the familiar monopoles,
it was not clear what role these solitons
should play in the quantum theory. Quantum mechanically, the black holes
and black $p$-branes are not stable due to Hawking radiation. This radiation
is absent for the extremely charged solutions, so we are lead to focus on them.
Unfortunately, there is an immediate problem. Due to the coupling to the
dilaton, the extremal limit of black holes and
black $p$-branes with a single type of charge is quite different from
the Reissner-Nordstr\"om solution. In string theory, the dilaton diverges
at the inner horizon causing this surface to become singular. 
As one approaches the
extremal limit, the event horizon approaches this ``inner horizon" and becomes
singular itself\footnote{Of course, once 
the curvature exceeds the string scale, one cannot usually trust the 
solution to the low energy equations. However, in some cases,
one can argue
that the low energy extremal solutions are in fact exact even when the
curvature diverges \cite{hots}.}.
To obtain the quantum states associated with a soliton, one usually 
identifies its
zero modes (or collective coordinates) and quantizes them. This is not
possible when the soliton is singular.

To resolve this problem, we need two ingredients. The first is supersymmetry.
In theories where the supersymmetry algebra has a central charge,
one can prove a general inequality
on the mass and charge of all perturbative states:  
$M\ge k Q$ where $k$ is a constant which can depend on the parameters
in the theory.  States which saturate this
bound are called BPS states, and have the important property that their
mass cannot receive quantum corrections. The number of BPS states is
essentially a topological invariant, and does not change if one
continuously changes the coupling constant, or other parameters in the theory.
(This is because BPS states form short representations of
the supersymmetry algebra, and the number of degrees of freedom
cannot change when the parameters are varied.)

The second ingredient is a better understanding of the charges
that exist in string theory. There are two basic types of gauge fields
called \NS (for Neveu-Schwarz) and \R  (for Ramond) and they can  carry
either electric or magnetic type charge. Both types of fields exist
in the low energy string action, and one can find solitons with each of
these charges. These fields differ in how the dilaton couples to them.
The charges are all quantized, and the difference in coupling leads
to a difference in the mass of an extremal soliton with one unit of charge.
In units with $l_s =1$, the mass of an extremal soliton with
one unit of electric \NS charge is of order one, the mass of a soliton
with one unit of magnetic
\NS charge is of order $1/g^2$, and with one unit of \R charge (either
electric or magnetic) is of order $1/g$. So when the string coupling $g$
is very small, the \R charged solitons and magnetic \NS charged solitons
are both very massive.

We now ask what is the gravitational field of these objects at weak coupling
when $g\rightarrow 0$. The  strength of this gravitational field is,
of course, determined by $GM$. The key point is that since Newton's constant
is proportional to $g^2$
$GM \rightarrow 0$ as $g\rightarrow 0$ for all $M < 1/g^2$. So the spacetime
associated with these solitons becomes flat as $g\rightarrow 0$ not only
for the electric \NS charged objects, but also for the \R charged ones. Since
the spacetime becomes flat, there should be a {\it nonsingular} description
of these states at weak coupling. What is this description?
For the electric \NS solitons, the
weak coupling description turns out to be just (some of) the usual perturbative
string states. This is because a fundamental string couples to the
\NS potential and can carry electric
\NS charge.
The usual perturbative states of the (super) string
satisfy $M\ge kQ$, and the BPS states with $M=kQ$ are the nonsingular
weak coupling description of the extreme electric \NS charged black holes.

At first sight, 
the identification of a perturbative string state with an extreme
black hole may sound absurd. After all, perturbative
string states, defined by quantizing a string in flat spacetime, represent
linearized fluctuations about the flat background. How can they be
equivalent to an extreme black hole? The answer, of course, is that 
the black hole
represents the result of including the gravitational backreaction to
the perturbative string state. As one increases the string coupling,
the mass does not change (since $M$ is independent of $g$ classically,
and supersymmetry forbids any quantum corrections). But the gravitational
field it produces becomes much stronger, and is described by a curved
spacetime with large curvature. The transition point is roughly the following:
If the nonlinear terms in the curved spacetime solution appear when the
curvature is less than $1/l_s^2$, then the curved spacetime description is
justified. If not, the flat spacetime description is  better.
The identification of massive BPS string states
with extremal black holes has been checked by explicit calculations of 
large impact parameter  scattering. Since scattering at large impact
parameter does not probe the nonlinear part of the geometry, 
the scattering of BPS string states computed in string
perturbation theory  should agree with the scattering of extremal
black holes computed
in a moduli space approximation. This is indeed the case \cite{khmy, cmp}.

What is the flat space description of extreme black holes with \R charge?
These cannot be the usual perturbative
string states because strings do not carry \R charge.
This is because 
strings do not couple to the \R potential but only to its field strength.
The appropriate flat space description of these objects was found just
last fall by Polchinski \cite{pol,pcj}.
Roughly speaking, one adds
to the usual closed string theory a certain class of open strings whose
ends are subject to Dirichlet boundary conditions. This has the effect
of confining them to live on certain surfaces - 
called D-branes. The D stands for Dirichlet boundary conditions, {\it not} 
the dimension of the brane. D-branes can have any dimension, and
one can show that they 
carry precisely one unit of the appropriate \R charge.  D-branes are 
dynamical objects and the open string
states describe their fluctuations. The important point is
that with this new description, the number of states with \R charge
can be counted, just like the usual perturbative string states.

With this background,
I can now describe the explanation of black hole entropy that has emerged
from string theory. The starting point is the observation
that while  black holes with a single
charge have extremal limits with singular horizons, black holes carrying several
different types of charge can have extremal limits with regular horizons 
\cite{kalo}.
When several gauge fields are nonzero, the dilaton can be `stabilized'
and remains finite on the horizon in the extremal limit.
This allows the  event horizon 
to remain a nonsingular surface with finite area just like
the Reissner-Nordstr\"om metric. In fact, the Reissner-Nordstr\"om metric
is itself a solution to string theory, but several gauge fields must be nonzero
and the charge parameter in the metric is  a function of the charges
of each of these gauge fields. To understand black hole entropy,
one views the solitons with a single charge as the elementary constituents
from which one builds up extremal black holes with nonzero area.
(Notice that for this interpretation, it is important that the solitons
with a single charge did not have finite area horizons.)
One starts at weak coupling with a BPS state with several \R charges $Q_i$
nonzero.  This can be viewed as a bound state of several D-branes.
As we increase the string coupling, the gravitational field
becomes stronger and the metric becomes an extreme black hole - now with
finite horizon area. Each BPS state with the same $Q_i$ at  weak coupling
is described by the same black hole at strong coupling. The key question
is: How many BPS states are there in the weakly coupled string theory with
the charges $Q_i$? For large $Q_i$, the answer turns out to be $e^{A/4G}$ where
$A$ is the area of the horizon that appears at strong coupling \cite{stva}! This
appears almost miraculous. Unlike other attempts to explain black hole
entropy, one does not start with a black hole with area $A$. There is no
surface of area $A$ at weak coupling. 
One has simply flat spacetime, and one counts
states with certain charges.
To the best of my knowledge, this is the first time 
that the Bekenstein-Hawking entropy
has been obtained by counting states in a quantum theory of gravity in more than
$2+1$ dimensions. (In $2+1$ dimensions, this has been achieved using the
Chern-Simons formulation of the theory \cite{car,carlip}.)

We have described the connection between black holes and
weakly coupled string states in units where the string length $l_s$ is fixed
as the string coupling $g$ is varied.
This is useful for seeing how the flat space description
arises. It turns out that in these units $A \propto g^2$
so that the entropy $A/4G$ 
is independent of the string coupling. This means that at weak coupling,
the horizon size is much smaller than
the string scale, and the horizon curvature is much larger than this scale.
Hence the  nonlinear terms in the metric are not reliable, and
the flat space description is appropriate.
As one increases the coupling, the horizon grows larger than the string
scale  and the black hole description becomes valid. 
Alternatively,  one can  keep the Planck length fixed as $g$ is varied.
The picture is then
different.  The black hole has a fixed size (independent of $g$) which
can be large in the limit of large charge. For small $g$, the configuration
of D-branes (which has a size of order the string scale)
is much larger than the  horizon radius. But as one increases
$g$ the size of the D-brane state shrinks. At a critical  value of the coupling
the string state  undergoes gravitational
collapse and falls inside the horizon. This is perhaps analogous to the 
maximum mass limits for white dwarfs or neutron stars\footnote{I thank
J. Polchinski for suggesting this viewpoint.}.

The above result was first obtained for a five
dimensional extreme black hole. Since then, a number of different
types of black holes have been examined and the agreement between the
number of string states and the Bekenstein-Hawking entropy has been shown
to hold in each case. 
These include five dimensional extreme black holes with rotation \cite{bmpv},
slightly
nonextremal five dimensional black holes \cite{cama,host,blmpsv},
four dimensional extreme black holes \cite{mast,jkm},
and slightly nonextremal four dimensional black holes \cite{hlm}.
(These calculations were facilitated by some earlier counting of D-brane states
\cite{senb,
vafa,dama}.)
The five dimensional
case was considered first since one only needs three nonzero charges to 
obtain an extreme black hole with regular horizon. In four dimensions,
one needs four nonzero charges. The counting of states for slightly nonextremal
black holes is possible since even though one must count non-BPS states,
one can argue that under certain conditions, quantum corrections remain 
negligible. It is important to keep in mind that it is not
just a single number which is being calculated in two different ways
with the same result.
For nonextremal four dimensional black holes,
the entropy is a function
of several parameters including the mass, and four different charges.
The number of string states  agrees with the black hole entropy 
as a function of all of these parameters.

\vskip 1cm
\noindent {\bf III. Examples}
\vskip 0.7cm

Many of the \NS charges in string theory
arise from compactifying one or more spacetime
directions. In addition to the usual Kaluza-Klein gauge field
which arises from the metric in the usual way, there is another gauge
field coming from the dimensional reduction of a three form $H_{\mu\nu\rho}$
which appears as a massless field in string theory. From the way that a
string couples to the metric and $H$, one can show that a string with 
momentum in a compact direction carries electric Kaluza-Klein charge,
while a string that winds around a compact direction carries electric
charge associated with $H$.
Consider  flat spacetime with one direction $x$
compactified with radius $R$. Examples of BPS states with one type of charge
are an unexcited string wrapped $m$ times around the circle  
 (with no momentum),
or an unexcited string with momentum $P_x = n/R$ (and no winding). 
These are both examples of electric
\NS charges since they are carried by a fundamental
string. For each value of the charge there is just one quantum BPS state
with that charge (apart from the degeneracy of the supersymmetry multiplet).
When both charges are nonzero, there are many BPS states,
since it turns out that 
the oscillators associated with either the left moving modes or right
moving modes (but not both) can be excited.
The extreme black hole with two charges still has a singular horizon, but
if one draws a surface where the curvature reaches the string scale 
(indicating the boundary where the metric can no longer be trusted),
then it turns out that its area correctly counts the number of BPS string
states with two charges (up to an overall factor of order one arising from
the fact that the location of the surface has not been precisely specified)
\cite{sen,peet}. One can interpret this roughly as resulting from the fact that
the number of BPS states with two charges only grows like $e^M$, which is
not fast enough for the horizon size to become larger than the string scale.

To obtain an extreme black hole with finite area, we need to consider
more than two charges. We will discuss the simplest example, which is
a five dimensional black hole with three charges. To understand the
quantum states associated with this black hole, it is
more convenient to consider a six dimensional black string. As discussed
in the introduction, the entropy of the black string will be the same as
the black hole obtained by dimensional reduction. (For a more detailed 
discussion of this example, see \cite{host, hms}.) Since string
theory requires ten dimensions, we assume that the remaining four
dimensions are compactified on a fixed torus of volume $(2\pi)^4 V$
which is constant. We will also assume that the dilaton is constant.
The only nontrivial field besides the metric is then the three form
$H_{\mu\nu\rho}$. For six dimensional solutions with a spacelike 
translational symmetry, $H$
can carry
both electric and magnetic charges which are proportional to the integral
of $*H$ and
$H$ over the asymptotic three sphere  in the space orthogonal to the symmetry
direction. 
These charges are quantized and take integer values which we will denote
by $Q_1$ and $Q_5$. (The reason for this notation will become clear shortly.)
The solution to the low energy string equations turns out to be \cite{ght}
\be\label{metric}
ds^2= -\left (1- {\rp  \over r^2}\right )dt^2+ 
\left (1-{\rmn \over r^2}\right )dx^2 
~~+\left (1-{\rp\over r^2} \right )
^{-1}\left (1-{\rmn \over r^2}\right )
^{-1}dr^2
+r^2d\Omega^2_3
\ee
This metric is similar to the five dimensional Reissner-Nordstr\"om solution.
There is an event horizon at $r=r_+$ and an inner horizon at $r=r_-$.
It is static, spherically symmetric, and translationally invariant along
the $x$ direction. The parameters $r_+$ and $r_-$ are related to the charges
by $Q_1 Q_5 =  r_+^2 r_-^2 V/g^2$.
The extremal limit corresponds to $r_+ = r_-\equiv r_0$.
If we periodically identify $x$ with period $2\pi R$, the 
extremal ADM energy is 
\be
E_0 = {2r_0^2 RV\over g^2}
\ee
where we have used the fact that the six dimensional Newton's constant
is $G = \pi^2 g^2 /2V$ in units with $\l_s=1$.
In this case, the event horizon has zero area.
However, unlike the case of
extreme solutions with a single charge (or two electric charges), the 
curvature does not diverge at the horizon in the extremal limit. 
The horizon area  vanishes
simply because the length in the $x$ direction shrinks to zero.
To obtain an extremal solution
with nonzero area, we can add momentum along the string. (This provides the
third charge upon dimensional reduction to five dimensions.) Since the extremal
solution is boost invariant, we cannot add momentum by boosting it. Instead,
we start with the nonextremal solution, apply a boost $t = \hat t
\cosh\sigma
 + \hat x \sinh\sigma, \ x =  \hat x \cosh\sigma + \hat t   \sinh\sigma $,
 and identify $\hat x$ with period $2\pi R$.
The ADM energy of these solutions is now
\be
E={RV \over 2 g^2}\left [2(\rp +\rmn)+\cosh{2\sigma}
(\rp-\rmn)\right ]
\ee
and the $x$ component of the ADM momentum is 
\be
P={RV\over 2g^2  }{\rm sinh}2\sigma \ (\rp-\rmn) .
\ee
The horizon area 
is now
\be\label{area}
A = 4\pi^3 r_+^2 R \cosh \sigma \sqrt{\rp - \rmn}
\ee
and the Hawking temperature is\footnote{This corrects a misprint in
\cite{host}.}
\be
T = {\sqrt{\rp - \rmn}\over 2\pi \rp \cosh\sigma}
\ee
The extremal limit is obtained by taking $r_- \rightarrow r_+$ keeping
$P$ fixed, which requires $\sigma\rightarrow \infty$. The resulting
solutions have energy
\be
E_{ext} = E_0 +P
\ee
Since the energy is increased by an amount equal to the momentum,
the effect of boosting and taking the extremal limit is to add a null vector
to the total energy-momentum. This can be viewed as adding ``right moving"
momentum only. In the extremal limit, the Bekenstein-Hawking entropy is
\be
S_{BH} = {A\over 4 G} = 2\pi \sqrt{r_0^4V PR\over g^2} = 2\pi \sqrt{Q_1 Q_5 PR}
\ee
We now assume that $P$ is quantized $P= n/R$, since this will be the
case for the quantum states we wish to count. So the
entropy is simply
\be\label{bhent}
S_{BH} = 2\pi \sqrt{Q_1 Q_5 n}
\ee
Notice that this expression depends only on the integer charges, and is
independent of both the size of the compact directions and the string coupling.
The dimensional reduction of this extremal six dimensional black string
with momentum yields a five dimensional extremal black hole with the same
entropy. 

So far we have just discussed  properties of a classical solution.
We now turn to counting states in string theory. The solution
we have described arises in all string theories\footnote{There have
traditionally been five (super) string theories which differ in  the gauge
groups and amount
of supersymmetry one has in ten dimensions. However recent work has shown that
these theories
are all related by various strong-weak coupling duality transformations.}
since they all include a \NS field $H$. However, the magnetic \NS charge
does not have a flat space description, which makes the counting of states
difficult. However, one string theory (called Type IIB) has another three
form which is a \R field.  The solution for the black string is identical, and
one can now try to count its states in a weak coupled description.
We start
with ten dimensional flat space and compactify four dimensions on a torus 
of volume $(2\pi)^4 V$ and one direction on a
circle of length $2\pi R$ which is much larger than the other four.
We now ask what objects carry the charges $Q_1$ and $Q_5$. It turns out
that a D-fivebrane which wraps once around the five torus carries charge
$Q_5=1$, and a 
D-string wrapped once around the circle with radius $R$ carries charge
$Q_1=1$. The black strings
discussed above are thus bound states of
$Q_5$ fivebranes and $Q_1$ strings. If we set $Q_5 =1$, then the number of
such bound states is easy to count.  From the six dimensional viewpoint
these bound states look like a string with certain degrees of freedom
which arise as follows. Although the $Q_1$ strings are bound to the
fivebrane, they are still free to move
around in the four internal directions. This yields
$4Q_1$ massless bosons together with their superpartners on the $1+1$
dimensional
effective field theory on the string. (The massive excitations of the D-branes
have masses proportional to $1/g$ and hence do not play a role at weak coupling.
Both the massless and massive
degrees of freedom are represented by open strings with ends on the
D-branes.) For $Q_5 >1$,  this generalizes to
$4Q_1Q_5$ massless bosons. Free fields in $1+1$ dimensions have
independent right and left moving modes.
BPS states with nonzero momentum correspond
to exciting only the right moving modes of these massless fields. So 
one can simply count
the number of states of $4Q_1Q_5$ bosonic fields (plus an equal number of
fermionic fields required by supersymmetry) on a circle of radius $R$
with total right moving
momentum $P=n/R$. In the limit
of large $n$, the answer is $e^S$ where
\be\label{stent}
S = 2\pi \sqrt{Q_1 Q_5 n}
\ee
in perfect agreement with the Bekenstein-Hawking entropy (\ref{bhent}).
We have assumed that one compact direction is
much larger than the other four in order to simplify the counting of states,
but this is not essential. Since the number of BPS states does not change
when the radii are continuously varied, one knows that the entropy 
is given by (\ref{stent}) for all values of $R$. When $R$ is small, it
is more appropriate to interpret the six dimensional black string as a
five dimensional black hole.
Notice that from
the five dimensional standpoint, the nonzero entropy for an extremal black
hole appears to indicate a highly degenerate ground state. However, from the
standpoint of the six dimensional string, one has a nondegenerate ground
state with $P=0$, and the horizon area becomes nonzero only when momentum 
is added. 

To describe the states of slightly nonextremal black holes, one
keeps exactly the same degrees of freedom as before, but now allows
them to carry both left and right moving momentum. These are no longer
BPS states, and the counting cannot be done for all values of $R$. However,
when $R$ is large, the modes carrying the momentum have very low energy
and one would expect interactions between left and right moving modes
to be negligible even as one turns up the string coupling. If one
considers states with left moving momenta $P_L = n_L/R$ and right moving
momenta $P_R = n_R/R$ then, since the total entropy of noninteracting
systems is additive, one obtains
\be \label{neent}
S = 2\pi \sqrt{Q_1 Q_5}(\sqrt n_R + \sqrt n_L)
\ee
in the limit of large $n_R,\ n_L$.
It may seem difficult for an expression like 
this to arise in the context of black hole
entropy since the area involves one square root, not the sum of two
square roots. Nevertheless, (\ref{neent}) also agrees with 
the Bekenstein-Hawking entropy, as can be seen as follows. 
To first order away from extremality, $r_\pm = r_0 \pm \epsilon$,
the horizon area (\ref{area}) becomes
\be\label{smarea}
A = 8\pi^3 r_0^2 R \cosh\sigma \sqrt{r_0 \epsilon}
\ee
Let $\delta  E$ denote the ADM energy of the
black string
above the zero momentum ground state i.e. $\delta  E = E-E_0$,
and define left and right
moving momenta by
\be
P_R = {n_R\over R} \equiv { 1 \over 2  }(\delta  E+P)\qquad 
P_L = {n_L\over R} \equiv { 1 \over  2 }(\delta  E-P)
\ee
so $P_R-P_L=P$. Then it turns out that $n_R \propto e^{2\sigma},\
n_L \propto e^{-2\sigma}$ and the Bekenstein-Hawking entropy obtained
from (\ref{smarea}) becomes
precisely (\ref{neent}). This expression is valid when both the energy density
$\delta E/R$ and momentum density $P/R$ are small. The integers
$n_R$ and $n_L$ can still be large since $R$ is large. So we end up with a
remarkably simply picture of  black string states in terms of free
fields on a circle which describe both extremal and near extremal
configurations.

\vskip 1cm
\noindent {\bf IV. Discussion}
\vskip 0.7cm

It is ironic that an extremal black hole was the first example 
for which it was
explicitly shown that the number of states is $e^{A/4G}$.
It was argued recently \cite{hhr} that extremal black holes should have 
zero entropy even if they have nonzero area. This was based on the
fact that the analytic continuation of an extreme black hole has
a different topology than the analytic continuation of a nonextreme
hole.  For a nonextreme black hole, regularity at the horizon fixes
the periodicity of the euclidean time coordinate, and the topology is
$R^2 \times S^2$.  For an extreme black hole, 
since the horizon is infinitely far away 
one can periodically identify the euclidean time direction with any period and
the topology is $S^1 \times R \times S^2$. 
In a semiclassical approximation, the entropy is related to the action
of the euclidean black hole solution by 
\be
S=\left ( \beta {d\over d\beta} -1\right )I_E
\ee
where $\beta$ is the periodicity of the euclidean time at infinity. The 
entropy is nonzero only if the action contains a contribution which is
not proportional to $\beta$. For stationary configurations in ordinary
field theory, $I_E = \beta H$ where $H$ is the hamiltonian, and the
entropy is zero. For a nonextreme black hole,  since the stationary
surfaces all intersect at the horizon, it is convenient to write the action
as the sum of a contribution from a small neighborhood of the horizon and 
the contribution from everything outside. This second term is indeed
proportional to $\beta$, but the first is not and yields the familiar
result $S=A/4G$. However an extreme black hole seems analogous to 
ordinary field theory. The stationary surfaces do not intersect, the action
is proportional to $\beta$, and the entropy would appear to be zero.

String theory offers a natural resolution of this apparent 
contradiction\footnote{The following argument was developed in discussions with
A. Sen.}.
Consider a Reissner-Nordstr\"om solution with large charge. The curvature of
the euclidean solution is everywhere small, so one expects that
string corrections should be negligible. While this shows that there is an
exact solution of this form, it is not stable. For any periodicity $\beta$
at infinity, the length of the euclidean time direction goes to zero
asymptotically as one approaches the horizon. This means that it
must eventually become of order the string scale. At this point, the
partition function diverges. One can interpret this divergence as
resulting from the fact that a winding
mode of the string which wraps around this circle becomes massless \cite{atwi}.
When
the circle becomes smaller than the string scale, this mode becomes
tachyonic. The original euclidean Reissner-Nordstr\"om metric corresponds
to a solution in which this mode has been strictly set to zero. This is
clearly unstable in the region where it becomes tachyonic. So one expects
that the stable description of the extreme black hole will resemble
the  Reissner-Nordstr\"om solution in the region where the euclidean time
circles are larger than the string scale, but differ significantly 
beyond this point. In particular, there is no reason to expect the topology
to remain $S^1\times R\times S^2$. Since the modification to the geometry
is likely to involve string scale curvature, it is difficult to describe
the stable solution explicitly.  However, two things are clear:
(1) Some modification of the solution must occur at the string scale
since this is directly analogous to the Hagadorn transition which has
been well studied for strings at finite temperature in flat spacetime.
(2) Since the modification always occurs at the string scale, it is likely to
give a contribution to the action which is independent of $\beta$ and hence
lead to a nonzero entropy. It would be nice to show explicitly that the
stable solution leads to $S=A/4G$, e.g. by showing that the stable solution
has the same topology as the nonextremal black hole.

We now consider the question of which states are being counted to 
represent the black hole entropy. For the case of extreme black holes,
one simply counts all BPS states in the weakly coupled 
string theory with the given charges. 
This is consistent with the
fact that the only solutions of the low energy field equations with mass
and charge saturating 
the inequality $M > kQ$ are extreme black holes. (This follows from a 
positive energy theorem.) Since there is no force between
these extremal black holes, there are static solutions describing
several extreme black holes, but for a fixed charge, the maximum 
entropy configuration is a single black hole. Similarly,
the maximum number of BPS states arise when
the D-branes are bound together, and hence localized
to about a string length in spacetime. 
For the nearly extremal
black holes, it is not true  that one counts all states in the weakly
coupled theory with the same mass and charges. For example,
we do not include states
obtained by adding  very low energy strings which are far from 
a BPS D-brane bound state. Instead, the states which are counted
are just non-BPS excitations of the D-branes. These are still localized
to about a string length in spacetime and form a black hole at strong 
coupling. They can be viewed as the weakly coupled states associated with
the near extremal black hole.

As we have discussed, string theory has provided  a statistical explanation of
the entropy of extremal and  nearly
extremal black holes. What are the prospects for extending these results
to black holes far from extremality, including the Schwarzschild solution?
First note that at finite string
coupling $g$, it is plausible that  essentially all 
perturbative string states with mass $M > 1/g^2 l_s$ (including uncharged
states) are black holes \cite{sus}.
The reason is similar to the old argument that any particle with mass
greater than the Planck mass should be a black hole since its Compton
wavelength is less than its Schwarzschild radius. In string theory
this argument must be modified since the metric is not really well defined
unless the curvature is less than the string scale $1/l_s^2$.
Since (in four dimensions) the
curvature at the horizon of a  black hole of mass $M$ is of order 
$1/(GM)^2$ and
Newton's constant
is related to $l_s$ by $G\sim g^2 l^2_s$, the metric is
well defined at the horizon in string theory provided $M > 1/g^2 l_s$ which
is $1/g$ times the Planck mass. One consequence of this
argument is that the most massive string states are essentially stable,
and decay only through semi-classical Hawking evaporation. One occasionally
hears  the statement that most massive string states decay very quickly to 
lighter string states, and only BPS states can be stable. This is based
on a perturbative calculation which is not valid at finite string coupling.
Gravitational
backreaction is strong and forms a black hole. 

However, we have already noted that this cannot be the explanation of
black hole entropy since the degeneracy of massive perturbative
string states only
grows like $e^M$ and not $e^{M^2}$. Clearly, one must also 
include solitonic states. From our experience with  extremal black holes, one
might expect 
that solitons provide the dominant contribution to the entropy.
 Consider the six dimensional black string discussed in the
previous section. We showed that the quantum states of the 
extremal and near extremal configurations could be understood in terms
of a collection of weakly interacting onebranes, fivebranes and strings
(carrying the momentum). It turns out that there is a sense in which 
the general nonextremal solution can also be viewed as composed of weakly
interacting objects which include these solitons as well as `anti-branes'
which  are just D-branes with the opposite orientation and carrying the 
opposite sign of the charge \cite{hms}. To see this, one first generalizes the 
solution slightly so that the volume of the internal four torus is no
longer constant. Dimensional reduction then yields a  five dimensional black
hole which depends
on six parameters: the mass, three charges, and the asymptotic value of 
two scalars (which represent the radius of the $S^1$ and volume of the
$T^4$ in ten dimensions). Alternatively, one can label the solution by
its  mass, three gauge charges and two `scalar charges'. One might think
that the black hole no-hair theorems would imply that the scalar charges
are determined by the gauge charges and are not independent. However,
if one normalizes the gauge charges to take integer values,  the scalar
charges depend on both these integers and the asymptotic values of the scalars.
One can then replace these six parameters 
by a number of branes, anti-branes and strings
($N_1,\ N_{\bar 1},\ N_5,\ N_{\bar 5},\ n_R, \ n_L$) by matching
the mass, three gauge charges and two scalar charges of
these noninteracting objects with that
of the black hole. In terms of these new
variables the Bekenstein-Hawking entropy takes the suggestive form \cite{hms}
\be
S_{BH}= 2 \pi( \sqrt{ N_1} + \sqrt{  N_{ \bar 1}})
( \sqrt{ N_5} + \sqrt{  N_{ \bar 5}})( \sqrt{ n_L} + \sqrt{ n_R})~.
\ee
This expression applies to all five dimensional black holes, even
the Schwarzschild solution. If one term in each factor vanishes, the black hole
is extremal and the entropy agrees with the number of states of this collection
of D-branes at weak coupling. If only one of the three factors has both terms
nonzero, this entropy can again be shown to correctly count the number of
states at weak coupling. What is missing is a general derivation of this
expression from counting states of branes, anti-branes and strings
when all six are present.
There is a similar expression for four dimensional black holes
with four different factors corresponding to the 
different solitons that carry  the four different charges \cite{hlm}.

Another open question concerns corrections to the Bekenstein-Hawking
entropy formula. This formula gives the leading contribution to the entropy for
large black holes. For smaller black holes one expects corrections arising from
one loop quantum effects, as well as corrections 
arising from higher order terms in the classical
string equations of motion. The latter contribution is likely to involve
integrals of local functions of the curvature over the horizon \cite{wald}.
These corrections should all be
reflected in the counting of quantum string states.

Now that the quantum states associated with black holes have been
identified, one can try to resolve the question of whether black holes
lose information when they radiate. Hawking has argued that if an extreme
black hole absorbs a small amount of energy to become slightly nonextremal,
and then radiates back to an extremal state, information is lost. In the
weak coupling D-brane
description, this corresponds to a low energy closed string state scattering
off a BPS configuration of D-branes. This scattering can be
computed in perturbation theory and is certainly unitary. However,  to resolve
the issue of information loss, one must do much more. As we have discussed,
as one turns down the string coupling, the horizon size becomes very small
compared to the string scale. Spacetime becomes flat, and the black hole
is effectively replaced by boundary
conditions on the D-brane. The region of spacetime behind the
horizon is not represented in this weak coupling description. It is hardly
surprising that information is not lost in a context where there is
no place for it to go! Although it may be possible to relate this
D-brane description to what is seen by observers who stay outside
a black hole (at strong coupling), it is clearly not sufficient to describe
what happens to observers who fall in.

In order to have a complete
understanding of the information puzzle, it is necessary to establish
stronger links between the flat space D-brane picture, and the
black hole picture.
One possible approach is the following. The extremal limit
of the six dimensional black string (\ref{metric})
has an unusual global structure.
This spacetime has a (degenerate) horizon, but is completely nonsingular 
\cite{ght}. 
The region of spacetime behind the horizon is isometric to the region in front.
If one starts with this solution and turns down the coupling, one obtains
two copies of Minkowski spacetime, each with a D-brane, but with these
D-branes identified\footnote{Actually, the maximally extended spacetime
contains an
infinite number of asymptotically flat regions.}.
In a sense, one represents the region behind the
horizon, and the other represents the region in front. The idea would be
to try to modify the usual D-brane perturbation rules so that strings entering
from one flat space, could exit into the other. This would correspond
to something falling through the horizon. Unfortunately, it is not yet clear
how this should be done. Another approach to strengthening the links between
the two pictures is to understand what happens to the D-branes when the
string coupling is increased. It seems unlikely that the D-branes can stay
on the event horizon - like all massive objects, they presumably fall in. 
At strong coupling, the quantum states are then localized within
a string length of the singularity. But what happens in the case of the
extremal black string when there is no singularity? Also, if the quantum
states are localized near the singularity, how are they connected with 
Hawking radiation? In the weakly coupled
D-brane picture, the natural analog of Hawking radiation is for two open
string states on the D-brane to combine, form a closed string, and leave
the D-brane. Can one follow  this
process as the string coupling is increased? Is it related 
to the familiar semiclassical calculations of black hole radiation
for large black holes?

Finally, the explanation of black hole entropy provided by string theory
needs to be understood at a deeper level. At the present time, each 
black hole must be checked by a separate calculation, and the agreement
between the number of string states and the Bekenstein-Hawking entropy
seems miraculous. There must be some unifying principle which explains
why these calculations are working. It is clear that string theory
has not yet revealed all of its secrets.

\vskip 1cm
\noindent {\bf  Acknowledgments}
\vskip 0.7cm
It is a pleasure to thank D. Marolf, J. Polchinski, and C. Rovelli for 
helpful conversations. I also wish to thank my collaborators,
D. Lowe, J. Maldacena,
and especially A. Strominger for extensive discussions which clarified my
understanding and led to 
some of the results described here.
This work was supported in part
by NSF Grant PHY95-07065.

\newpage
\renewcommand{\refname}
\end{document}